\documentclass{elsart4}

\usepackage{graphicx}

\usepackage{amssymb,amsbsy,amsmath}

\newcommand{\op}[1]{%
    \fontdimen12\textfont3=2pt\fontdimen12\scriptfont3=1.4pt%
    \!\null\mathop{\vphantom{#1}\smash{#1}}\limits_{\sim}\null\!}
\newcommand{\xref}[1]{\protect\ref{#1}}
\newcommand{\figref}[1]{Fig.~\protect\ref{#1}}
\newcommand{\fmref}[1]{(\protect\ref{#1})}
\def\bra#1{\langle \, {#1} \, | \,}
\def\ket#1{\, | \, {#1} \, \rangle}

\newtheorem{k-rule}{k-rule}

\begin{document}
\begin{frontmatter}

\title{Solitary waves on finite-size antiferromagnetic quantum Heisenberg spin rings}


\author{J\"urgen Schnack\corauthref{cor1}}
\author{Pavlo Shchelokovskyy}
\address{Universit\"at Osnabr\"uck, Fachbereich Physik,
D-49069 Osnabr\"uck, Germany}
\corauth[cor1]{phone: ++49 541 969-2695; fax: -12695; Email: jschnack@uos.de}

\begin{abstract}
  Motivated by the successful synthesis of several molecular
  quantum spin rings we are investigating whether such systems
  can host magnetic solitary waves. The small size of these spin
  systems forbids the application of a classical or continuum
  limit. We therefore investigate whether the time-dependent
  Schr\"odinger equation itself permits solitary waves. Example
  solutions are obtained via complete diagonalization of the
  underlying Heisenberg Hamiltonian.
\end{abstract}

\end{frontmatter}

\section{Introduction}
\label{sec-1}

Magnetic solitons are detected in many magnetic systems due to
their special influence on magnetic
observables \cite{KLH:PRL99,Essler:PRB99,ANI:PRL00,LKG:PB00,NAA:PB01}.
From a theoretical point of view magnetic solitons are solutions
of non-linear differential equations, e.~g. of the cubic
Schr\"odinger equation \cite{Mat88,MiS:AP91,Mik:CSF95}, which
result from classical approximations of the respective quantum
spin problem. The cubic Schr\"odinger equation for instance is
obtained if the spins are replaced by a classical spin
density \cite{Mat88}.

Due to the success of coordination chemistry one can nowadays
realize finite size quantum spin rings in the form of magnetic
ring molecules. Such wheels -- Fe$_6$, Fe$_{10}$, and Cr$_8$
rings as the most prominent examples -- are almost perfect
Heisenberg spin rings with a single isotropic antiferromagnetic
exchange parameter and weak uniaxial
anisotropy \cite{TDP:JACS94,ACC:ICA00,WKS:IO01,VSG:CEJ02}.

The aim of the present article is to discuss whether solitary
waves can exist on such spin rings and if they do, how they look
like. The finite size and the resulting discreteness of the
energy spectrum forbid any classical or continuum limit. We
therefore investigate, whether the ordinary \emph{linear}
time-dependent Schr\"odinger equation allows for solitary waves.
Solitary excitations in quantum spin chains of Ising or
sine-Gordon type have already been discussed \cite{MiS:AP91}.
Nevertheless, such soliton solutions are approximate in the
sense that a kind of Holstein-Primakov series expansion is
applied, and the results are accurate only for anisotropies
large compared to the spin-spin coupling \cite{MiS:AP91}. This
article deals with antiferromagnetic Heisenberg chains without
anisotropy, where such derivations cannot be applied. Starting
from the time-dependent Schr\"odinger equation therefore
automatically addresses the questions how soliton solutions
might be approached starting from a full quantum treatment -- a
question that to the best of our knowledge is not yet
answered \cite{Mik:PC}.

In order to apply the concept of solitons or solitary waves to
the linear Schr\"odinger equation some redefinitions are
necessary. The first redefinition concerns the term soliton
itself. It is mostly used for domain-wall like solitons which
are of topological character. It is also applied to localized
deviations of the magnetization or energy distribution (envelope
solitons), here one distinguishes between bright and dark
solitons. We will generalize this class of objects and speak
only of solitary waves in the following. We call a state
$\ket{\Psi_s}$ solitary wave if there exists a time $\tau$ for
which the time evolution equals (up to a global phase) the shift
by one site on the spin ring. This means that solitary waves
travel with permanent shape. The property that two solitons
scatter into soliton states cannot be used as a definition in
the context of the Schr\"odinger equation since it is trivially
fulfilled for a linear differential equation.

A nontrivial question concerns useful observables in order to
visualize solitary waves $\ket{\Psi_s}$. The expectation value
of the local operator $\op{s}_z(i)$, which reflects a local
magnetization is used classically and meaningful also as a
quantum mechanical expectation value. Then nontrivial, i.~e.
non-constant magnetization distributions
$\bra{\Psi_s}\op{s}_z(i)\ket{\Psi_s}$ arise only for those
solitary waves which possess components with a common total
magnetic quantum number $M$, because matrix elements of
$\op{s}_z(i)$ between states of different total magnetic quantum
number vanish.  The energy density which in classical
calculations is also used to picture solitons, could quantum
mechanically be defined as \linebreak $\op{\vec{s}}(i) \cdot
\op{\vec{s}}(i+1)$ starting from a Heisenberg Hamiltonian,
compare Eq.~\fmref{E-2-1}. It turns out that this observable is
not so useful since it is featureless in most cases because the
off-diagonal elements for energy eigenstates belonging to
different total spin $S$ are zero.

The article is organized as follows. In Sec.~\ref{sec-2} we
shortly introduce the Heisenberg model and the used notation,
while in Sec.~\ref{sec-3} solitary waves are defined. We will
then discuss the construction and stability of (approximate)
solitary waves in Sec.~\ref{sec-7}. A number of example spin
systems will be investigated in Sec.~\ref{sec-4}. The article
closes with conclusions in Sec.~\ref{sec-6}.

\section{Heisenberg model}
\label{sec-2}

The Hamilton operator of the Heisenberg model with
antiferromagnetic, isotropic nearest neighbor interaction between
spins of equal spin quantum number $s$ is given by
\begin{eqnarray}
\label{E-2-1}
\op{H}
&\equiv&
2\,
\sum_{i=1}^N\;
\op{\vec{s}}(i) \cdot \op{\vec{s}}(i+1)
\ ,\quad N+1\equiv 1
\ .
\end{eqnarray}
$\op{H}$ is invariant under cyclic shifts generated by the shift
operator $\op{T}$. $\op{T}$ is defined by its action on the
product basis $\ket{\vec{m}}$
\begin{eqnarray}
\label{E-2-3}
\op{T}\,
\ket{m_1, \dots, m_N}
\equiv
\ket{m_N, m_1, \dots, m_{N-1}}
\ ,
\end{eqnarray}
where the product basis is constructed from single-particle
eigenstates of all $\op{s}_z(i)$
\begin{eqnarray}
\label{E-2-2}
\op{s}_z(i)\,
\ket{m_1, \dots, m_N}
=
m_i\,
\ket{m_1, \dots, m_N}
\ .
\end{eqnarray}
The shift quantum number $k=0,\dots, N-1$ modulo $N$ labels the
eigenvalues of $\op{T}$ which are the $N$-th roots of unity
\begin{eqnarray}
\label{E-2-4}
z
=
\exp\left\{
-i \frac{2\pi k}{N}
\right\}
\ .
\end{eqnarray}
$k$ is related to the ``crystal momentum'' via $p=2\pi k/N$.

Altogether $\op{H}$, $\op{T}$, the square $\op{\vec{S}}^2$, and
the $z$-component $\op{S}_z$ of the total spin are four
commuting operators.

\section{Solitary waves}
\label{sec-3}

We call a state $\ket{\Psi_s}$ solitary wave if there exists a
time $\tau$ for which the time evolution equals (up to a global
phase) the shift by one site on the spin ring either to the left
or to the right, i.~e.
\begin{eqnarray}
\label{E-3-1}
\op{U}(\tau)\,\ket{\Psi_s}
&=&
e^{-i\Phi_0}
\;
\op{T}^{\pm 1}\,\ket{\Psi_s}
\ .
\end{eqnarray}
Decomposing $\ket{\Psi_s}$ into simultaneous eigenstates
$\ket{\Psi_{\nu}}$ of $\op{H}$ and $\op{T}$,
\begin{eqnarray}
\label{E-3-2}
\ket{\Psi_s}
&=&
\sum_{\nu\in I_s}\;
c_{\nu} \ket{\Psi_{\nu}}
\ ,
\end{eqnarray}
yields the following relation 
\begin{eqnarray}
\label{E-3-3}
\sum_{\nu\in I_s}\;
e^{-i\frac{E_{\nu}\tau}{\hbar}}
c_{\nu} \ket{\Psi_{\nu}}
=
e^{-i\Phi_0}
\sum_{\mu\in I_s}\;
e^{\mp i\frac{2 \pi k_{\mu}}{N}}
c_{\mu} \ket{\Psi_{\mu}}
\ .
\end{eqnarray}
The set $I_s$ contains the indices of those eigenstates which
contribute to the solitary wave.
Therefore,
\begin{eqnarray}
\label{E-3-4}
\frac{E_{\mu} \tau}{\hbar}
&=&
\pm\frac{2 \pi k_{\mu}}{N}
+ 2\pi m_\mu
+ \Phi_0
\qquad\forall \mu\in I_s
\\
&&
\text{with}\ 
m_\mu \in\mathbb{Z}
\nonumber
\ .
\end{eqnarray}
Equation \fmref{E-3-4} means nothing else than that solitary
waves are formed from such simultaneous eigenstates
$\ket{\Psi_{\nu}}$ of $\op{H}$ and $\op{T}$ which fulfill a
(generalized) linear dispersion relation. 

Since our definition \eqref{E-3-4} is rather general it
comprises several solutions. Some of them are rather trivial
waves whereas others indeed do possess soliton character:
\begin{itemize}
\item Single eigenstates $\ket{\Psi_{\nu}}$ of the Hamiltonian
  fulfill the definition, but they are of course stationary and
  they possess a constant magnetization distribution. One would
  not call these states solitary waves or solitons.
\item Superpositions of two eigenstates $\ket{\Psi_{\mu}}$ and
  $\ket{\Psi_{\nu}}$ with different shift quantum numbers
  $k_{\mu}$ and $k_{\nu}$ are also solutions of definition
  \fmref{E-3-4} since two points are always on a line in the
  $E$-$k$-plane. It is clear that such a state cannot be well
  localized because of the very limited number of momentum
  components. Nevertheless, these states move around the spin
  ring with permanent shape.

  An example of such solitary waves are superpositions
  consisting of ground state and first excited state which have
  already been investigated under a different
  focus \cite{HML:EPJB02}. The characteristic time $\tau$ to move
  by one site is related to the frequency of the coherent spin
  quantum dynamics discussed for these special
  superpositions \cite{HML:EPJB02,MeL:PB03}.
\item The existence of more than two eigenstates with linear
  dispersion relation cannot be proven for Heisenberg models in
  contrast to the Ising model \cite{MiS:AP91}. In fact it is very
  likely that such a combination does not exist at all for a
  small system. Therefore, superpositions of three or more
  eigenstates will -- hopefully only slightly -- deviate from
  the linear dispersion relation and therefore describe (slowly)
  decaying solitary waves. This will be explained in detail
  below.
\item The character of the solitary waves assembled according to
  definition \eqref{E-3-4} can be different. As
  section~\ref{sec-4} will show some of them are localized
  magnetization distributions (envelope solitons) which could be
  denoted as bright or dark solitary waves others are of
  topological nature as for example in the case of
  superpositions of the ground state and the first excited state
  on an odd membered spin ring.  Nevertheless, the distinction
  is not sharp. As shown in Ref.~\cite{Lak:PL77} envelope
  solitons can be of topological nature when looked at from a
  different perspective.
\end{itemize}

\section{Construction and stability of solitary waves}
\label{sec-7}

When constructing solitary waves we focus on two principles:
\begin{itemize}
\item Energetically low-lying solitons can be formed by
  superimposing the ground state with low-lying excited states.
  For this purpose it is advantageous that for antiferromagnetic
  rings the lowest energy eigenvalues depend on momentum
  approximately via a sine function. Therefore, for large
  systems, where the sine function can be approximated by its
  argument, an almost linear dependence between $k_{\mu}$ and
  $E_{\mu}$ is expected.
\item General solitons can be formed by superimposing states
  with arbitrary pairs $(k_{\mu},E_{\mu})$ as long as they
  fulfill \eqref{E-3-4}. For this purpose two states are
  selected which define a straight line in $E-k$--space. Then a
  search algorithm is used to find other states whose energy and
  momentum quantum numbers constitute a point on (or in close
  vicinity) of the line. The search extends not only to the
  first Brilluoin zone, but also to higher zones.
\end{itemize}

It is clear that in a finite system the linear dispersion
relation \eqref{E-3-4} might only be fulfilled approximately
which leads to decaying solitary waves. That the time evolution
must be recurrent in a Hilbert space of finite dimension is only
of principle interest since the recurrence time would be very
long.

A good measure of the stability of a solitary wave is the
overlap of the time-evolved state and the shifted one. For
perfect stability these two states coincide. Let's assume that
the wave travels in the same direction as the shift operator
acts, then we define the stability measure as
\begin{eqnarray}
\label{E-7-1}
\eta(\tau)
&=&
\bra{\Psi_s}\op{T}^{-1}\,\op{U}(\tau)\,\ket{\Psi_s}
\ .
\end{eqnarray}
$\eta(\tau)$ is a complex number, and for dispersionless
movement it's absolute value equals one.  Decomposing the
approximate solitary wave $\ket{\Psi_s}$ into simultaneous
eigenstates $\ket{\Psi_{\nu}}$ of $\op{H}$ and $\op{T}$ yields
\begin{eqnarray}
\label{E-7-2}
\eta(\tau)
&=&
\sum_{\nu\in I_s}\;
|c_{\nu}|^2\,
e^{-\frac{i \tau}{\hbar}
\left\{
E_{\nu}-
\frac{2 \pi \hbar k_{\nu}}{N \tau}
\right\}
}
\\
&=&
\sum_{\nu\in I_s}\;
|c_{\nu}|^2\,
e^{-i \frac{\delta_{\nu} \tau}{\hbar}}
\nonumber
\ ,
\end{eqnarray}
where $\delta_{\nu}$ is the energy mismatch of energy level
$\nu$. In order to consider a motion across $n$ sites the
exponent has to be multiplied by $n$.

As a simple example we want to consider an approximate solitary
wave which consists of three eigenstates as for instance in the
example of \figref{F-4}. The first two levels define the linear
dependence, and the third level deviates by $\delta_3$ from that
dependence. Then, taking the normalization condition into
account, the overlap turns out to be
\begin{eqnarray}
\label{E-7-3}
\eta(\tau)
&=&
1-|c_3|^2 + |c_3|^2 e^{-i \frac{\delta_{3} \tau}{\hbar}}
\ .
\end{eqnarray}
Thus we face a decaying overlap which later oscillates back to
one. For the specific example of \figref{F-4} we can estimate
how much time this will take. The characteristic time $\tau$ is
given by
\begin{eqnarray}
\label{E-7-4}
\frac{\hbar}{\tau}
&=&
\frac{\Delta E}{\Delta k}
\frac{N}{2 \pi}
\ ,
\end{eqnarray}
which is approximately 3 in the given units, compare
\figref{F-1}. The energy mismatch is of the order of $10^{-2}$.
In order to have a deviation of the exponential in \eqref{E-7-3}
of 1~\% from unity the argument has to grow up to $0.1$, which
means that the solitary wave has to move across about 30 sites. Or
in other words, the approximate three state solitary wave of
\figref{F-4} has to circumvent the spin ring three times in
order to produce a deviation of the overlap with the original
state of about one percent. This means that such approximate
solitary waves are rather stable. Nevertheless, with increasing
energy mismatch the situation will be worse.

\section{Special solutions for spin rings}
\label{sec-4}

In this section we discuss some typical examples for magnetic
solitary waves on finite spin rings. We have chosen examples
which are as large as numerically possible, i.e. $N=11, 12, 14$
and $s=1/2$, because a too small size yields only a very limited
number of eigenstates with different shift (i.e. momentum)
quantum numbers that can be superimposed to form the solitary
wave.  Then, the solitary waves obviously cannot be of such
perfect shape and localization as known from continuous models.
Nevertheless, our general ideas are applicable as well to
systems with spin quantum numbers $s>1/2$, which in future
investigations might be accessible by means of Density Matrix
Renormalization Group (DMRG) techniques \cite{Sch:RMP05}.

\begin{figure}[!ht]
\begin{center}
\includegraphics[clip,width=60mm]{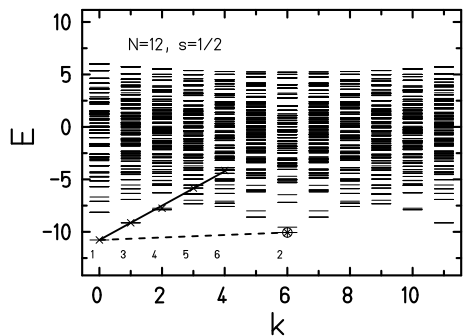}
\vspace*{1mm}
\caption[]{Energy spectrum of a antiferromagnetic spin ring with
  $N=12$ and $s=1/2$: The solid as well as the dashed line
  connect states which are superimposed to form a solitary wave.
  The numbers label these states.}
\label{F-1}
\end{center}
\end{figure}

Figure~\xref{F-1} shows a typical spectrum for an
antiferromagnetically coupled Heisenberg ring. The ground state
of such a ring is a singlet with $k=0$ for the special case of
$N=12$. The first excited state is a triplet with $k=6$, compare
the theorems of Lieb, Schultz, and
Mattis \cite{LSM:AP61,LiM:JMP62,Wal:PRB02} for quantum numbers of ring
systems with an even number of sites and
Ref.~\cite{BHS:PRB03} for quantum
numbers of ring systems with an odd number of sites.

\begin{figure}[!ht]
\begin{center}
\includegraphics[clip,width=60mm]{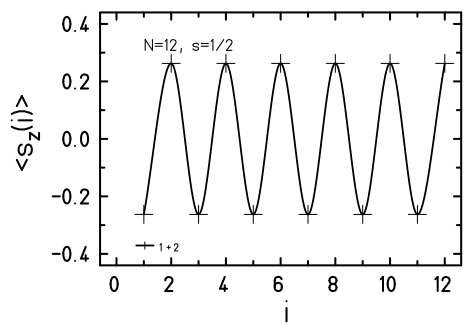}
\vspace*{1mm}
\caption[]{Solitary wave for $N=12$ and $s=1/2$ consisting of
  ground state and first excited state with $M=0$ (crosses). The
  curve is drawn as a guide for the eye. In the course of time
  such a wave moves around the spin ring without dispersion. The
  local magnetization distribution is the quantum expression of
  the classical N\'eel state.}
\label{F-2}
\end{center}
\end{figure}

Two basic types of low-energy solitary waves can be formed. The
first one is a superposition of the ground state and the triplet
state with $M=0$ (dashed line in \figref{F-1}). If added up the
result is a sinusoidal magnetization function -- sometimes also
called spin-density wave -- which is depicted in \figref{F-2}. A
subtraction leads to an inverted local magnetization. In the
course of time these distributions just move along the spin ring
without dispersion. This dynamics is equivalent to an
oscillation of the local magnetization, which was intensely
discussed under the aspect of tunneling of the N\'eel vector
\cite{HML:EPJB02,Waldmann:EPL02,MeL:PB03}.

\begin{figure}[!ht]
\begin{center}
\includegraphics[clip,width=60mm]{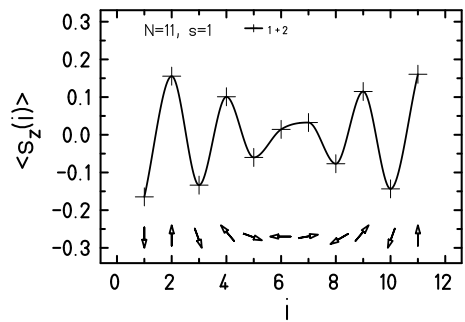}
\vspace*{1mm}
\caption[]{Solitary wave for $N=11$ and $s=1$ consisting of
  ground state ($k=0$) and first excited state ($k=5$) with
  $M=0$ (crosses). The curve is drawn as a guide for the eye.
  The local magnetization distribution is the quantum expression
  of a topological solitary wave, where the Ne\'el up-down
  sequence is broken and continued with a displacement of one
  site. This state propagates without dispersion.}
\label{F-3}
\end{center}
\end{figure}

Although these states look rather unspectacular the same
procedure applied to rings of an odd number of spins leads to
topological solitary waves. The reason is that due the periodic
boundary condition the antiferromagnetic order is frustrated in
such cases.  Figure \xref{F-3} shows the case of eleven spins
with $s=1$.  Here the ground state has $S=0$ and $k=0$ and the
first excited one, which is sixfold degenerate, has $S=1$ and
$k=5$ as well as $k=6$. The superposition which is depicted in
\figref{F-3} consists of the non-degenerate ground state and the
$(M=0,k=5)$-component of the first excited state.  One realizes
that this state describes a displacement of the classical Ne\'el
order by one site. As sketched by the arrows the Ne\'el order
starts at the l.h.s. as a sequence where up spins are located at
even sites and down spins at odd sites.  This sequence is
separated by a ``domain wall region'' from the sequence at the
r.h.s. where up spins are located at odd sites and down spins at
even sites. Such a behavior of the local magnetization can
equally well be pictured as a M\"obius strip, compare
Ref.~\cite{CGS:JMMM04}, although the system discussed in that
reference does not possess translational symmetry.

\begin{figure}[!ht]
\begin{center}
\includegraphics[clip,width=60mm]{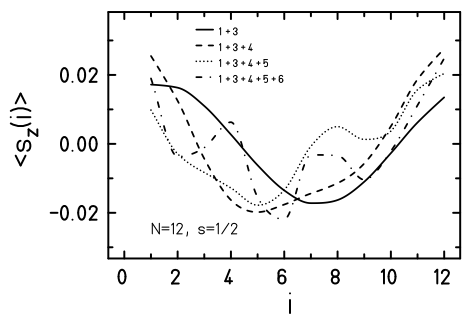}
\vspace*{1mm}
\caption[]{Several solitary waves depending on contributing
  eigenstates of $\op{H}$ and $\op{T}$ with $M=0$. All states
  contribute with the same weight in this presentation. All
  states with more than two components disperse slowly according
  to \fmref{E-7-4}. We estimate that the overlap with the
  initial state decreases by about 1~\% after two cycles around
  the ring.}
\label{F-4}
\end{center}
\end{figure}

The second basic type of low-energy solitary waves is built of
states along the solid line in \figref{F-1}. Adding up two to
five components with the same amplitude yields the magnetization
distributions shown in \figref{F-4}. These solitary waves are
similar to two domain walls on a larger ring system which
separate a spin-up region from a spin-down region (Two domain
walls are needed because the ring has an even number of sites.).
For a small system such as the investigated one the size of the two
domain walls is similar to the size of the whole system. The
solitary waves in \figref{F-4} with more than two components are
not stable because they fulfill the linear dispersion relation
\fmref{E-3-4} only approximately, compare Sec.~\xref{sec-7}.

\begin{figure}[!ht]
\begin{center}
\includegraphics[clip,width=60mm]{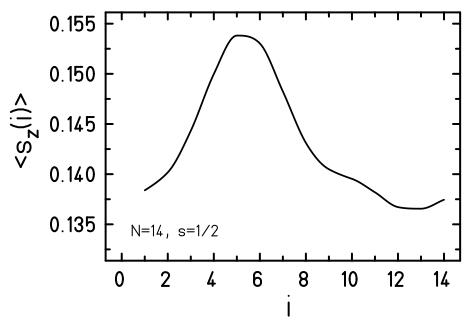}
\vspace*{1mm}
\caption[]{Solitary wave built of six components for $N=14$,
  $s=1/2$, and $M=2$ \cite{Shchelokovskyy:04}. The deviation of
  the energy levels from a strict linear behavior is less than
  $10^{-4}$, which means that this state disperses very slowly,
  i.e., it can perform many cycles around the ring without
  noticable deviation from the initial shape.}
\label{F-5}
\end{center}
\end{figure}

The above discussed low-energy solitary waves have a magnetic
quantum number $M=0$ and are accessible in low-temperature
experiments. Of course, solitary waves can be constructed also
for larger magnetic quantum numbers $M$ although it might not be
easy to excite or detect them.  Figure \xref{F-5} presents an
example of a rather well localized solitary wave in a subspace
with non-zero magnetic quantum number \cite{Shchelokovskyy:04}.

\section{Conclusions}
\label{sec-6}

Summarizing, we have presented a framework to discuss solutions
of the time-dependent Schr\"odinger equation which are of
permanent shape, i.e. travel without dispersion. Besides
solutions which have a close affinity to solitary waves our
rather general definition \eqref{E-3-4} also comprises solutions
which are simpler waves, which nevertheless propagate without
dispersion.  In general one can say, that due to the two reasons
that the investigated systems are small and that the Heisenberg
Hamiltonian \fmref{E-2-1} does not possess any anisotropy the
observed solitary waves are rather broad.

The special case of a superposition of the ground state and the
first excited state of a spin ring should be well observable at
low temperatures for instance by Nuclear Magnetic Resonance
(NMR). For rings with an even number of sites this is currently
aimed at \cite{HML:EPJB02,MeL:PB03}. Rings with an odd number of
sites are difficult to produce due to steric arrangement
problems of the extended ligands. First attempts resulted in
rings where one paramagnetic ion is different from the
others \cite{CGS:JMMM04}. The prospects of interesting features
due to frustration nevertheless fuel future efforts to
synthesize odd rings.

\begin{figure}[!ht]
\begin{center}
\includegraphics[clip,width=60mm]{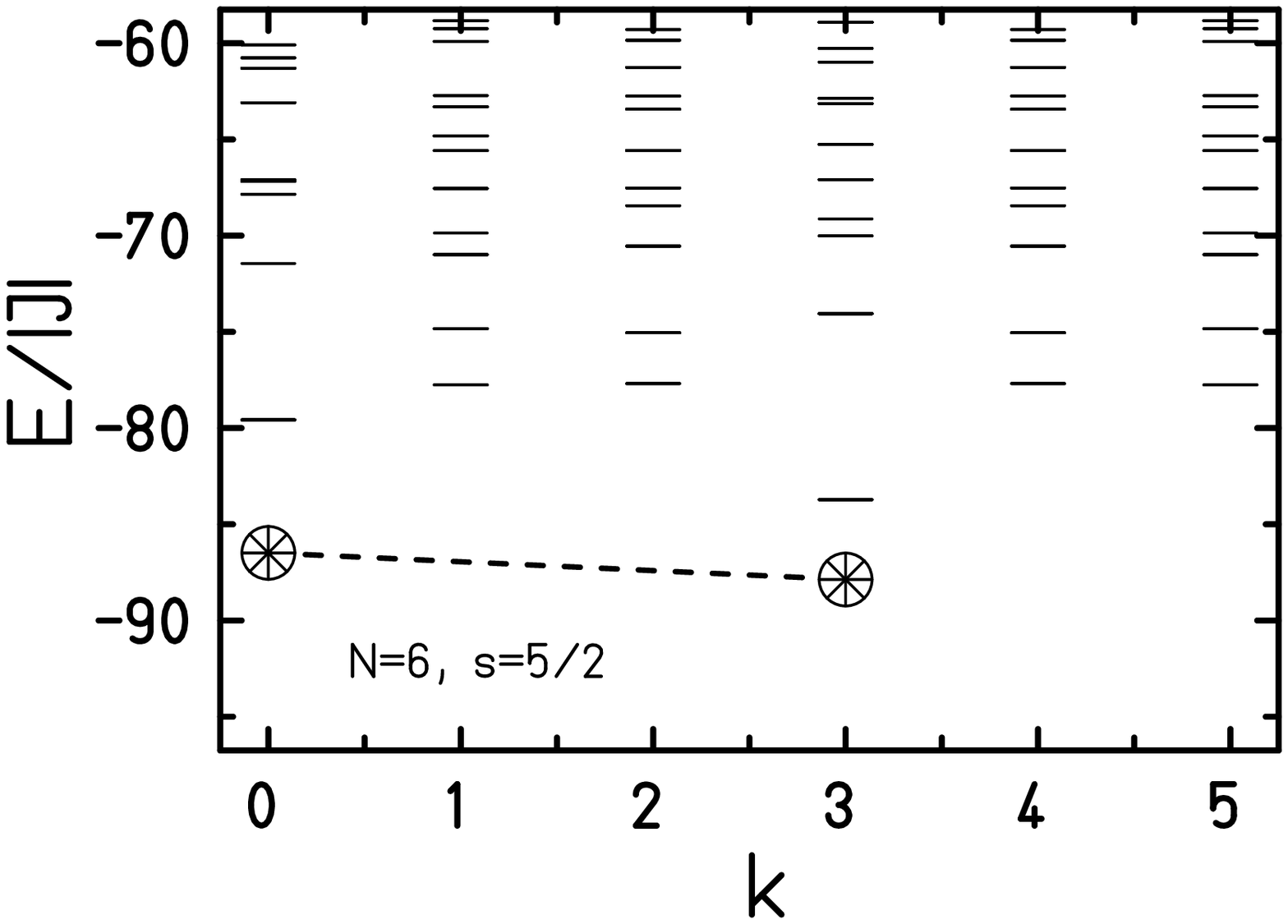}

\includegraphics[clip,width=60mm]{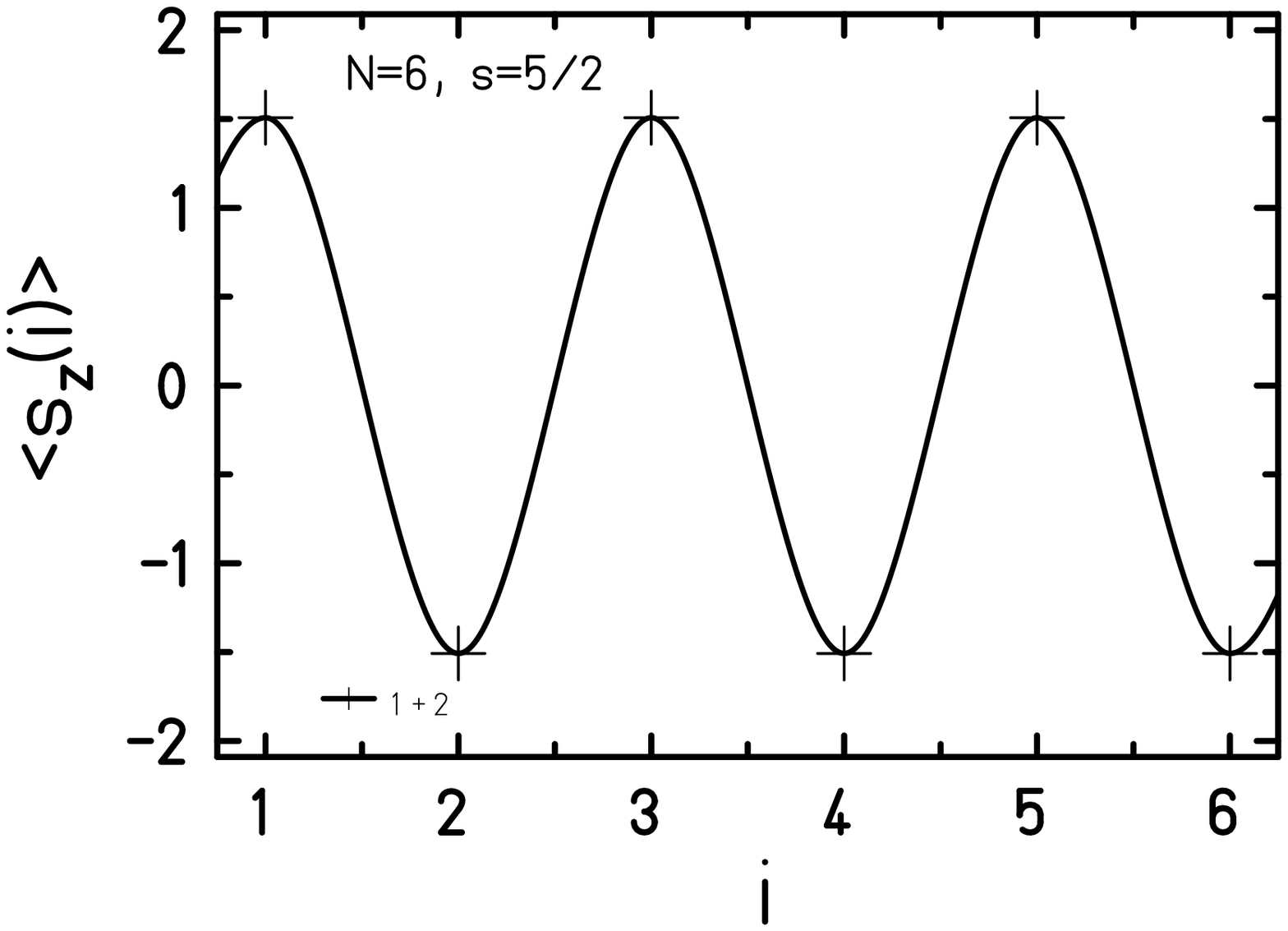}
\vspace*{1mm}
\caption[]{Top panel: Low-lying energy spectrum of a
  antiferromagnetic spin ring with $N=6$ and $s=5/2$: The dashed
  line connects the ground state and the first excited state. 

  Bottom panel: Solitary wave for $N=6$ and $s=5/2$ consisting
  of ground state ($k=3$) and first excited state ($k=0$) with
  $M=0$ (crosses). The curve is drawn as a guide for the eye.}
\label{F-6}
\end{center}
\end{figure}

Another important aspect is the role of anisotropic terms in the
Hamiltonian. Such terms are relevant for the magnetic properties
of several magnetic molecules, e.g. for ferric wheels comprising
six iron ions of spin $s=5/2$
\cite{CCF:CEJ96,WSK:IO99,Waldmann:EPL02}.  Easy axis anisotropy
would for instance amplify the magnetization oscillations, which
arise when ground state and first excited state are
superimposed. Figure \xref{F-6} shows in its bottom panel such
oscillations again, this time for a hexanuclear iron ring. One
notices that the local magnetization reaches at most an
amplitude of 1.6 due to strong quantum fluctuations in the
Heisenberg model. Increasing easy axis anisotropy would result in
more Ising-like behavior with increased local magnetizations. It
will be the subject of further studies how that in general would
influence solitary waves, but it is already clear that under
such conditions solitary waves would be more localized
\cite{LKL:PRB02}.

\section*{Acknowledgments}

We would like to thank Jochen Gemmer, Felix Homann, Uwe Sander,
Heinz-J\"urgen Schmidt, and Hans-J\"urgen Mikeska for helpful
discussions.


\end{document}